\documentclass[a4paper,11pt]{article}
\usepackage{jcappub} % for details on the use of the package, please see the JINST-author-manual
\usepackage{lineno}
\usepackage{subfig}
\usepackage[capitalise]{cleveref}
\usepackage{graphicx}% Include figure files
\usepackage[dvipsnames]{xcolor}
\usepackage{dcolumn}% Align table columns on decimal point
\usepackage{bm}% bold math
\usepackage{orcidlink}

\title{Forecasts and Simulations for Relativistic Corrections to the Sunyaev-Zeldovich Effect}

\author[a,b]{L. Kuhn\,\orcidlink{0009-0005-0766-5026}}
\author[b,c]{Z. Li\,\orcidlink{0000-0002-0309-9750}}
\author[d,e]{William~R.~Coulton\,\orcidlink{0000-0002-1297-3673}}
\affiliation[a]{Department of Physics \& Astronomy, University of British Columbia,\\ 6224 Agricultural Road, Vancouver BC, V6T 1Z1, Canada}
\affiliation[b]{Canadian Institute for Theoretical Astrophysics,\\ University of Toronto, 60 St. George ST., Toronto ON, M5S 3H8, Canada}
\affiliation[c]{Berkeley Center for Cosmological Physics,\\ University of California, Berkeley, CA 94720, United States}
\affiliation[d]{Kavli Institute for Cosmology Cambridge,\\ Madingley Road, Cambridge CB3 0HA, UK}
\affiliation[e]{DAMTP, Centre for Mathematical Sciences,\\ University of Cambridge, Wilberforce Road, Cambridge CB3 OWA, UK}

\emailAdd{lkuhn@phas.ubc.ca}

\abstract{The Sunyaev-Zeldovich (SZ) effect is a window into the astrophysical processes of galaxy clusters, and relativistic corrections (the ``rSZ'') promise to provide a global census of the gas feedback within clusters.
Upcoming wide-field millimeter-wave surveys such as the Simons Observatory (SO), Fred Young Submillimeter Telescope, and CMB-S4 will make increasingly precise measurements of the SZ effect and its relativistic corrections.
We present simulated full-sky maps of the rSZ effect and a fast code to generate it, for use in the development of analysis techniques and pipelines.
As part of the \textsc{websky} simulation suite, our mock observations have semi-realistic cross-correlations with other large-scale structure tracers, offering insights into the formation and evolution of galaxy clusters and large-scale structure. As a demonstration of this, we examine what an SO-like experiment can learn from the rSZ effect. We find that high significance detections will be possible, provided that the instrumental systematics are under control, and that the evolution of cluster temperatures with mass and redshift can be probed in a manner complementary to X-ray measurements.}

\begin{document}
\maketitle
\flushbottom

\section{\label{sec:level1}Introduction}

Secondary anisotropies in the CMB, such as the Sunyaev-Zeldovich (SZ) effects \citep{1969Ap&SS...4..301Z,1970Ap&SS...7...20S, 2008RPPh...71f6902A, 1999PhR...310...97B},
provide a unique view into the astrophysics of galaxy groups and clusters. Two key effects are the thermal and kinetic SZ (tSZ and kSZ) effects, which arise from Compton scattering of CMB photons by electrons with large thermal and bulk velocities respectively, and they trace out the momentum and pressure of cosmic gas.
 Ground-based experiments such as the Atacama Cosmology Telescope (ACT) \cite{2024PhRvD.109f3530C}, the South Pole Telescope (SPT) \cite{2015ApJS..216...27B}, and the Simons Observatory (SO) \cite{2019JCAP...02..056A} as well as space-based observations such as the Planck satellite \cite{2016A&A...594A..22P,2024PhRvD.109b3528M,2023MNRAS.526.5682C}, 
have precisely measured these SZ effects across large areas of the sky \citep[see, e.g.,][]{2016A&A...594A..22P,2016A&A...586A.140P,2020PhRvD.102b3534M,Coulton_2023,2024PhRvD.109b3528M}. These measurements are already starting to present challenges to leading cosmological galaxy formation models \citep{Schaan_2021,Hadzhiyska_2024}.

The SZ effects are usually analyzed with the non-relativistic approximation, where the thermal and bulk velocities of the scattering electrons are assumed to be small; however, cosmic velocities and gas temperatures can be so large that these approximations break down. The corrections to the non-relativistic treatment are known as relativistic corrections (or relativistic SZ, rSZ, effects). The relativistic corrections alter both the spectral shape of tSZ and kSZ effects and encode extra astrophysical and cosmological information \citep{2020PhRvL.125k1301C,2024arXiv240300909D}. Ignoring these effects can induce potential biases in analyses of the SZ effects \cite{Remazeilles_2024} and discards a new source of astrophysical information.

The detection and characterization of the rSZ effect pose significant challenges due to its relatively weak signal compared to the tSZ and kSZ effects. Several works have constrained these rSZ signals for both single clusters observations \citep{2013MNRAS.430.3054C,2012MNRAS.424L..49P,2002ApJ...573L..69H,2022ApJ...932...55B} and large samples of clusters \citep{hurier16, erler18, Remazeilles_2024,coulton_2024b}.
Upcoming experiments with improved sensitivity and higher angular resolution, such as the Simons Observatory \cite{2019JCAP...02..056A}, the Fred Young Submillimeter Telescope (FYST or previously CCAT-prime) \cite{2023ApJS..264....7C} and CMB-S4 \citep{2016arXiv161002743A}, will enable precision measurements of the rSZ effect in the next decade. These efforts will enable the thermodynamics of galaxy clusters to be measured across cosmic time and provide new insights into the gastrophysical processes that shape cluster environments \citep{2022PhRvD.105h3505T,2024arXiv240300909D}. 

In recent years, hydrodynamical cosmological simulations have become increasingly vital for modeling the distribution and properties of hot, ionized gas in galaxy clusters, helping to provide crucial insights into the rSZ. These simulations incorporate the complex baryonic physics governing the intracluster medium (ICM), including cooling, star formation, and various feedback processes such as active galactic nucleus (AGN) feedback and supernova-driven winds \citep{Voit2005}. AGN feedback, in particular, plays a significant role in regulating the thermal state of the ICM by heating the gas and preventing excessive cooling, which in turn influences observable properties such as the SZ effect \citep{2012ApJ...758...75B}. Current hydrodynamical simulations, such as those from the \texttt{IllustrisTNG} \citep{Springel2018}, \texttt{BAHAMAS} \citep{2017MNRAS.465.2936M}, and \texttt{Magneticum} \citep{2016MNRAS.463.1797D} projects, have achieved considerable success in reproducing observed properties of galaxy clusters. However, there remain uncertainties in the modeling of feedback processes and the precise calibration of these simulations, the process of tuning the parameters governing the feedback models to match observations \citep{Kravtsov2012}. Calibration is typically performed by matching simulated profiles of cluster observables, such as temperature and density, with observational data to ensure the simulated clusters exhibit realistic properties \citep{LeBrun2014}. Additionally, scaling relations between cluster properties (e.g., mass, temperature, and SZ flux) are used to link the simulations with observed cluster samples and to refine feedback models \citep{Pratt2009}. These scaling relations provide important constraints that help improve the physical realism of hydrodynamical simulations and their predictions for the SZ effect. Recently Refs. \citep{Lee_2020,2022MNRAS.517.5303L,Kay_2024} showed that rSZ temperature measurements are a sensitive probe of a cluster's thermal history.

Large suites of cosmological simulations are a key part of modern cosmological analyses: they are used to validate pipelines, test analysis choices, develop new summary statistics and train novel machine learning methods \citep[e.g.][]{2014JCAP...03..024O,2014ApJ...786...13V,2021PhRvD.104l3521H,2024ApJ...966..138M,2024MNRAS.533..423F}.
Unfortunately hydrodynamical simulations  are too expensive to use directly for simulating sky maps for large-area surveys \citep[however; see][for work in that direction]{2021ApJ...915...71V,2023MNRAS.526.4978S}. Thus, the leading approach is to generate large-scale, dark matter only N-body simulations and to then augment these to account for the missing baryonic physics.
This approach is accurate as the key effects of baryonic, astrophysical processes occur on small scales, whilst the large-scale matter distribution is accurately modeled with dark matter only simulations. A broad range of method exist to augment dark matter simulations including ``painting" profiles that have been fit from hydrodynamical simulations, analytical and phenomenological models that adjust the simulation particle position and properties, and machine learning approaches \citep{2012ApJ...758...75B,Battaglia_2016,2022MNRAS.517..420L,2022MLS&T...3c5002T,2019MNRAS.487L..24T,2023MNRAS.519.2069O}. 
Large-scale N-body simulations have been used to produce realizations of the extragalactic sky for a wide range of millimeter-wave observables \cite{Sehgal_2010, Han_2021, Stein_2020, Omori_2022, Bayer_2024}, including the tSZ, kSZ, Cosmic Infrared Background (CIB), and CMB lensing convergence — and the focus of this work: the relativistic SZ.
These simulated maps enable methods and pipeline development, and facilitate cross-survey analyses.

This paper describes a method create mock rSZ observations by painting profiles onto dark matter halo light cones and one demonstration of how such mocks can be used to explore future measurement prospects. Our approach is highly complementary to the only other work on this topic, Ref.\citep{Sehgal_2010} who used a particle based painting method rather than the halo based method used here. 
The structure of this paper is as follows. In Section \ref{sec:methods}, we discuss how we calculate the tSZ, kSZ and rSZ signals considering different mass-to-temperature scaling ratios. In \ref{sec:res_and_disc} we discuss the conversion of these 1D signals to 2D full sky maps which have implications for forecasting and cross-correlation analysis. In \ref{sec:conclusion}, we summarize our findings.

Throughout this paper, we use the cosmology defined by $\Lambda$CDM with $\Omega_{c,0}=0.2589$, $\Omega_{b,0}=0.0486$, and $h=0.6774\;\mathrm{km\;}s^{-1}\mathrm{Mpc}^{-1}$ as in \citep{2020JCAP...10..012S}

\section{\label{sec:methods}The Sunyaev Zeldovich effects}

The Sunyaev Zeldovich effects arise from Compton scattering of CMB photons off free electrons in Universe. Typically these effects are broken into two contributions the kinetic SZ effect and the thermal SZ effect. The kSZ effect arises from Compton scattering off electrons with large bulk velocities and the tSZ effect arises from Compton scattering off electrons with large thermal velocities. In this section we briefly review the non-relativistic treatments of the tSZ and kSZ effects, before discussing the full relativistic treatment of the SZ effects.
\subsection{Non-relativistic Thermal SZ effect}
Most commonly the tSZ effects is  modelled and analyzed in the non-relativistic approximation. This assumes that the electron's thermal velocities are non-relativistic, i.e. the electron temperature, $T_e$, is $T_e k_B\ll m_e c^2$. In this regime the tSZ
is described simply by the line-of-sight integration of the electron pressure \citep{1969Ap&SS...4..301Z,1970Ap&SS...7...20S},
\begin{equation}
   \Delta I(\mathbf{n})= f(X)y=f(X)\frac{\sigma_\mathrm{T}}{m_\mathrm{e}c^2}\int P_\mathrm{e}(l)\mathrm{d}l,
\end{equation}
\label{eqn:comp_y}
where $y$ is the Comptom-$y$ parameter, $f(X)=\left[X(\mathrm{e}^X+1)/(\mathrm{e}^X-1)-4\right]\left.\frac{\mathrm{d}B}{\mathrm{d}T}\right\vert_{T=T_\mathrm{CMB}}$ is the classical tSZ spectrum as a function of frequency-dependent dimensionless $X = h\nu/k_\mathrm{B}T_\mathrm{CMB}$, $\left.\frac{\mathrm{d}B}{\mathrm{d}T}\right\vert_{T=T_\mathrm{CMB}}$ is the derivative of the CMB blackbody, $P_\mathrm{e}(l)$ is the line of sight electron pressure, $m_e$ is the mass of the electron, $\sigma_T$ is the Thomson cross section and $c$ is the speed of light. For a fully ionized medium, the electron pressure can be expressed in terms of the thermal pressure $P_\mathrm{th}= P_\mathrm{e}(5X_\mathrm{H} + 3)/2(X_\mathrm{H} + 1) = 1.932P_\mathrm{e}$, where $X_\mathrm{H} = 0.76$ is the primordial hydrogen mass fraction \cite{2012ApJ...758...75B}. Battaglia et al. 2014 \citep{2012ApJ...758...75B} provide an empirical fitting function for the average thermal pressure profiles allowing for the following parameterization of the normalized thermal pressure $\bar{P}_\mathrm{th}$:
\begin{equation}
    \bar{P}_\mathrm{fit} = \frac{P_0(x/x_c)^\gamma}{\left[1 +(x/x_c)^\alpha\right]^{\beta}},
\end{equation}
\label{eqn:pfit}
where $x= r/R_\Delta$ where $r$ is the 3D distance from the cluster center and $R_\Delta$ is the spherical-overdensity radius, $P_0$, $x_c$, and $\beta$ are mass and redshift dependent constants determined using equation 11 and table 1 in \cite{2012ApJ...758...75B}, and $\alpha = 1.0$, $\gamma=-0.3$ are fixed constants. Taking the product of the normalized pressure profile with the self-similar amplitude for the pressure defined as $P_\Delta = GM_\Delta\Delta\rho_\mathrm{cr}(z)f_b/(2R_\Delta)$, the final thermal pressure can be written as:%
\begin{equation}
    P_\mathrm{th} =\frac{P_0(x/x_c)^\gamma}{\left[1 +(x/x_c)^\alpha\right]^{\beta}}\times \frac{GM_\Delta\Delta\rho_\mathrm{cr}(z)f_\mathrm{b}}{2R_\Delta}
\end{equation}
where $\rho_\mathrm{cr} = \frac{3H(z)^2}{8\pi G}$, $f_b=\Omega_b/\Omega_m$, $M_\Delta$ is the spherical-overdensity mass, and $\Delta$ is a threshold set to 200.

\subsection{Non-relativistic kinetic SZ effect}
The kSZ effect is often analyzed in the non-relativistic limit, where the cluster velocity, $\mathbf{v}$, is non-relativistic, i.e. $v/c\ll1$. In this regime the kSZ anisotropies are given by 
\begin{align}
    \Delta I(\mathbf{n},\nu) = -\mathcal{G}(x)\int \mathrm{d}l \sigma_T n_e(\mathbf{n},l)\frac{v\mu}{c},
\end{align}
where $n_e$ is the electron density, $\mathcal{G}(x)=\left.\frac{\mathrm{d}B}{\mathrm{d}T}\right\vert_{T=T_\mathrm{CMB}}$ and $\mu= \mathbf{\hat{v}}\cdot\mathbf{n}$. Ignoring internal cluster motions, the kSZ effect from a single cluster is given by
\begin{align}
    \Delta I(\mathbf{n},\nu) = -\mathcal{G}(x)\tau\frac{v\mu}{c},
\end{align}
where $\tau$ is the cluster optical depth.
Ref. \citep{Battaglia_2016} fit a normalized gas density profile, ${\rho_\mathrm{gas}}=\rho_\mathrm{crit}(z)\bar{\rho}_\mathrm{fit}$, as 
\begin{align}
    \bar{\rho}_\mathrm{fit} = \rho_0 \left( x/\tilde{x}_c \right)^{\tilde{\gamma}}\left[1+(x/\tilde{x}_c)^{\tilde{\alpha}} \right]^{-\tilde{\beta}}
\end{align}
where  $\rho_\mathrm{crit}(z)$ is the critical density, and $\tilde{x}_c$, $\tilde{\alpha}$, $\tilde{\beta}$, and $\tilde{\gamma}$ characterize the generalized NFW profile used to describe the gas density and are functions of the cluster mass and redshift. The gas density can then be related to the electron density via $n_e = \frac{x_e X_H}{\mu m_p}$ where $x_e$ is the ionization fraction, $m_p$ is the proton mass, $X_H$ is the primordial hydrogen mass fraction and  $\mu$ is the mean molecular weight for an ionized medium of primordial abundance.

\subsection{\label{sec:level2}The Relativistic SZ Effect}

\begin{figure}
\begin{center}
\includegraphics[width=0.7\textwidth]{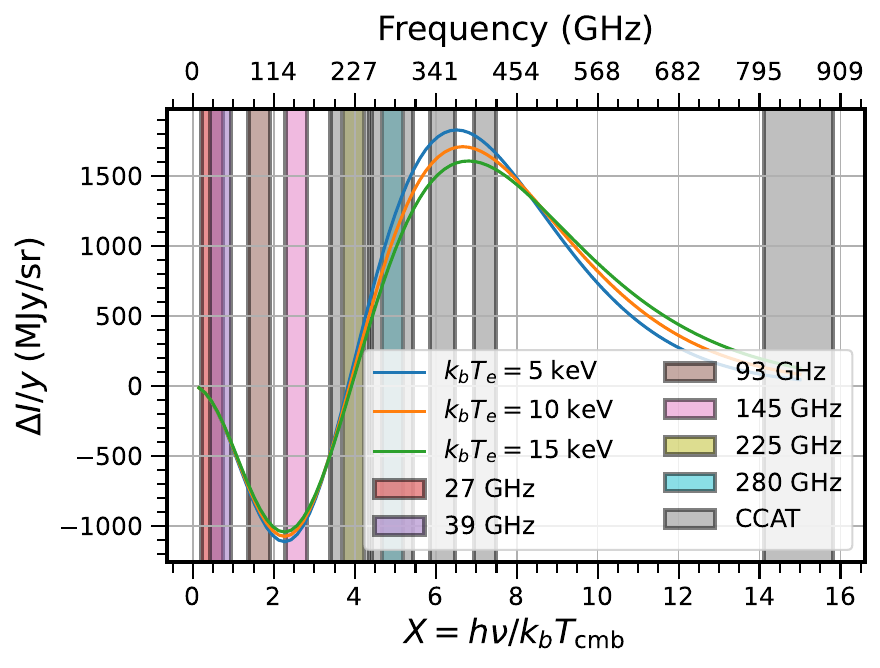}
\end{center}
\caption{The spatially independent intensity change resulting from rSZ profiles with different temperatures, including the tSZ equivalent (k$_B$T$_e=0$). We see the largest differences in these profiles near the peak frequency ($\approx$ 350 GHz). Note, as electron temperature increases, the profile is dampened and its peak frequency is shifted to larger values. The coloured bands here represent the frequency bands of Simmons Observatory, and the grey bands represent the upcoming CCAT/FYST.
\label{fig:rsz_temp}}
\end{figure}

Cluster temperatures can be hot enough, or their velocities large enough, that the non-relativistic approximations start to become inaccurate. For example, the temperatures of massive clusters can exceed $10\,$keV! Relativistic corrections describe corrections to the non-relativistic tSZ and kSZ effects described above, and the relativistic SZ effects (rSZ) describe the full signals accounting for the non-relativistic terms and the corrections. 

The relativistic SZ effects, to second order in $\beta_c=v/c$, are
\begin{align}\label{eq:rSZ_full}
   & \Delta I(x,\theta_e,\beta_c,\mu)  \nonumber\\& = y \left[ \mathcal{Y}(\theta_e,x)+\beta_c\mu  \mathcal{J}(\theta_e,x) +\beta_c^2P_2(\mu) \mathcal{L}(\theta_e,x)\right] \nonumber \\
   &+\tau\left[ \beta_c \mu \mathcal{G}(x)+\beta_c^2 P_2(\mu)\mathcal{K}(x)\right]
\end{align}
where the first term  $y\mathcal{Y}(\theta_e,x)$ is the relativistic equivalent of the thermal SZ effect, $\beta_c\mu \mathcal{J}(\theta_e,x) $  and $\beta_c^2P_2(\mu)\mathcal{L}(\theta_e,x)$ are mixed kSZ-tSZ terms,  $\beta_c \mu \mathcal{G}(x)$ is the non-relativistic kSZ term and $\beta_c^2 P_2(\mu)\mathcal{K}(x)$ is the relativistic correction to the kSZ effect. The expressions for these frequency-dependent functions are written out explicitly in Ref.~\cite{2006NCimB.121..487N}. Note: when writing this equation, we made a simplifying assumption that the temperature varies sufficiently slowly that the frequency functions can be evaluated at a single temperature for each cluster. This approximation dramatically speeds up our calculation and corresponds to a quasi-isothermal assumption (quasi as temperature variations still contribute to the Compton-$y$ parameter). From the weighted-pivot temperature viewpoint (as discussed in e.g. \citep{2020MNRAS.494.5734R}), this corresponds to ignoring the of the variance and higher order moments of the density weighted gas temperature and an assumption that the pivot temperature is the same for each term. For the precision of upcoming surveys we expect this to be a reasonable approximation and we will investigate relaxing this in future work. Cluster velocities are $\mathcal{O}(10^{-3})$ so higher order velocity terms are typically negligible. As discussed in e.g. \citep{2005A&A...434..811C,2013MNRAS.430.3054C} our motion also induces a distortion to the observed SZ effects. Whilst this is an interesting signal, especially in light of radio and quasar dipole measurements \citep{2020MNRAS.494.5734R}, we do not consider this here as this work focuses on extragalactic signals.

Whilst the perturbative expressions are available for both the tSZ and kSZ \citep{1998ApJ...499....1C,1998ApJ...502....7I,2000ApJ...536...31N,2006NCimB.121..487N,2012MNRAS.426..510C}, the tSZ expansion is only accurate at low temperatures and low frequencies. To enable modelling of the signal for any cluster temperature and the range of frequencies probed by SO and FYST we instead use \textsc{szpack} to evaluate the SZ signals
\citep{2012MNRAS.426..510C,2013MNRAS.430.3054C}. This code allows us to obtain precise SZ response functions for all temperatures and cluster velocities. This allows us to include mixed temperature and kinetic terms, which could be detected in the near future \citep[e.g.
][]{2020PhRvL.125k1301C}. We use the \texttt{COMBO} integration mode of \textsc{szpack}, which directly uses two different expansions to enable accurate and fast computations for a broad range of velocities ($v/c<$0.01) and temperatures ($0<T_e<65$\,keV. We refer the reader to Refs. \citep{2012MNRAS.426..510C,2013MNRAS.430.3054C} for details on the implementation.

We plot the spatially independent intensity change resulting from relativistic corrections to the tSZ profiles with different electron temperatures in Figure~\ref{fig:rsz_temp}. We observe the most substantial differences in the profiles near the peak frequency ($\approx$ 350 GHz). We note that as electron temperature increases, the profile is dampened and its peak frequency is shifted to larger values. The coloured bands here represent the frequency bands of Simmons Observatory.

\subsection{\label{sec:mass_temp}Mass-Temperature Scaling Relationships}

When adding relativistic corrections to the SZ effect, we also now must consider not just the cluster pressure and optical depth, but also the temperature of electrons along the line of sight of our integration. To do so, we make use of known galaxy cluster mass-to-electron temperature scaling relationships. 

In order to obtain information on the electron temperature of a cluster, we use the self-similar cluster formation prediction  \citep{1986MNRAS.222..323K,2008ApJ...672..752W}
\begin{align}
    T_\mathrm{self-sim}=\frac{\mu m_p}{2k_B}\frac{G M_\mathrm{vir}}{R_\mathrm{vir}},
\end{align}
where $\mu=0.6$ is  mean molecular weight, $m_p$ is the proton mass, $G$ is Newton's constant, $M_\mathrm{vir}$ is the virial mass and $R_\mathrm{vir}$ is the virial radius.

\begin{figure}
\begin{center}
\includegraphics[width=0.7\textwidth]{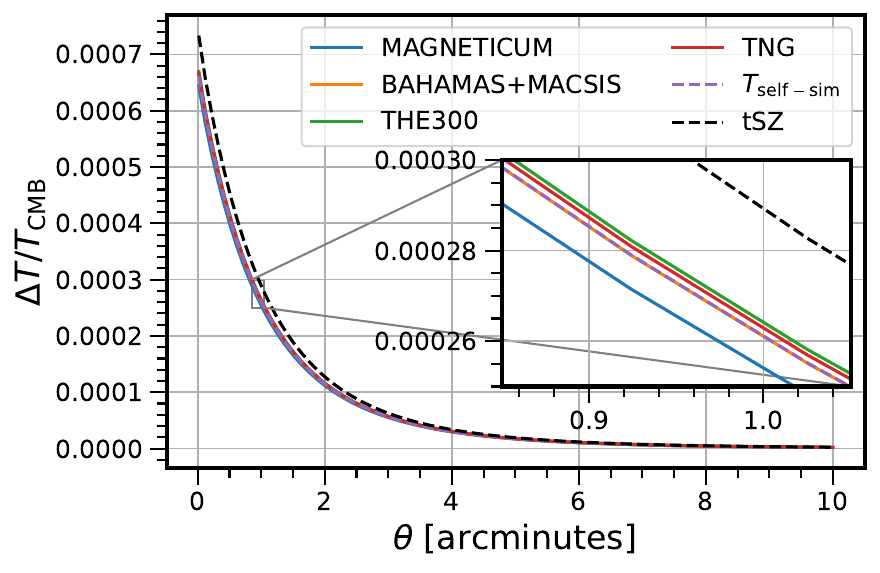} \\
\end{center}
\caption{
Comparison of the rSZ signal from a single massive cluster as a function of distance from the cluster center, in arcminutes,  using different mass-to-temperature scaling relationships. The mass-to-temperature scalings are obtained from fits to clusters from the \texttt{BAHAMAS}$+$\texttt{MACSIS}, \texttt{IllustrisTNG}, \texttt{MAGNETICUM}, and \texttt{THE THREE HUNDRED PROJECT} simulations. Inset provides a zoomed-in view for clarity. These profiles were generated for a frequency of $400$ GHz, a mass of $1.79\times10^{15} \mathrm{M}_\odot$, and $z = 0.4$. For comparison, the non relativistic tSZ effect, and the rSZ with a self-similar temperature are shown as dashed black and purple lines respectively.}
\label{fig:t_scale}
\end{figure}

Gas temperatures are not expected to perfectly follow the self-similar prediction --clusters are heated by AGN and supernova deviating their temperatures from self-similar predictions, see e.g. Ref. \citep{Churazov_2015}. This means that cluster temperatures can be an important probe of the history of cluster. To see this, we compare against cluster temperatures from a set of cosmological hydrodynamical simulations. Specifically, we use Ref. \cite{2022MNRAS.517.5303L} to test the differences between this relation and temperature-mass-redshift relations measured in the \texttt{BAHAMAS}$+$\texttt{MACSIS}, \texttt{IllustrisTNG}, \texttt{MAGNETICUM}, and \texttt{THE THREE HUNDRED PROJECT} simulations. We use the fits of the average rSZ temperature provided in table B9. Our comparison results are shown in Figure~\ref{fig:t_scale}.

We see significant differences in the predicted cluster signal depending on how AGN feedback is implemented. This demonstrates the value of precision cluster temperature measurements. For the rest of this work we consider the self-similar temperature relation.

\section{\label{sec:res_and_disc}Painting methodology and sky maps}

In this section we outline the method used to produce simulated maps, Sec. \ref{sec:maps}, and discuss the key features seen when we apply this method to the \textsc{Websky} simulations, Sec. \ref{sec:forecastImp}. 

The Websky simulations \citep{2020JCAP...10..012S, 2022JCAP...08..029L} are a set of extragalactic millimeter-wave mock observations, generated based on an approximate N-body gravity solver and phenomenological models of sources like dusty galaxies from the Cosmic Infrared Background, radio galaxies, as well as the thermal and kinetic Sunyaev-Zeldovich effects. They have been used extensively for forecasting and analysis validation for ground-based CMB telescopes.

\subsection{Painting methodology}\label{sec:maps}

We extend the \texttt{XGPaint} package \cite{XGPaintASCL} to produce rSZ maps. \texttt{XGPaint} generates mock rSZ observations by combining a dark-matter halo catalog with fitting formulae for the 3D gas thermodynamic profiles (currently assumed to be spherically symmetric). The code integrates the relevant 3D gas properties along the line of sight to obtain a projected profile (e.g. of the gas density and gas pressure); these are the key ingredients needed to evaluate \cref{eq:rSZ_full}. High resolution catalogs can contain 10s of billions of dark matter halos and so to enable fast synthesis \texttt{XGPaint} relies on a precomputed interpolation table of the projected profiles. In fact \cref{eq:rSZ_full} is implemented with two interpolators: one interpolator for $y$ and a second for $\tau$ that are then combined with their respective frequency dependencies. Once an input halo catalog and model type are specified, the signal can be painted on the sky employing the user's choice of pixelization schemes (i.e. CAR or HEALPix).

\subsection{\label{sec:forecastImp}Websky sky maps}

\begin{figure*}
\begin{center}
\includegraphics[width=0.98\textwidth]{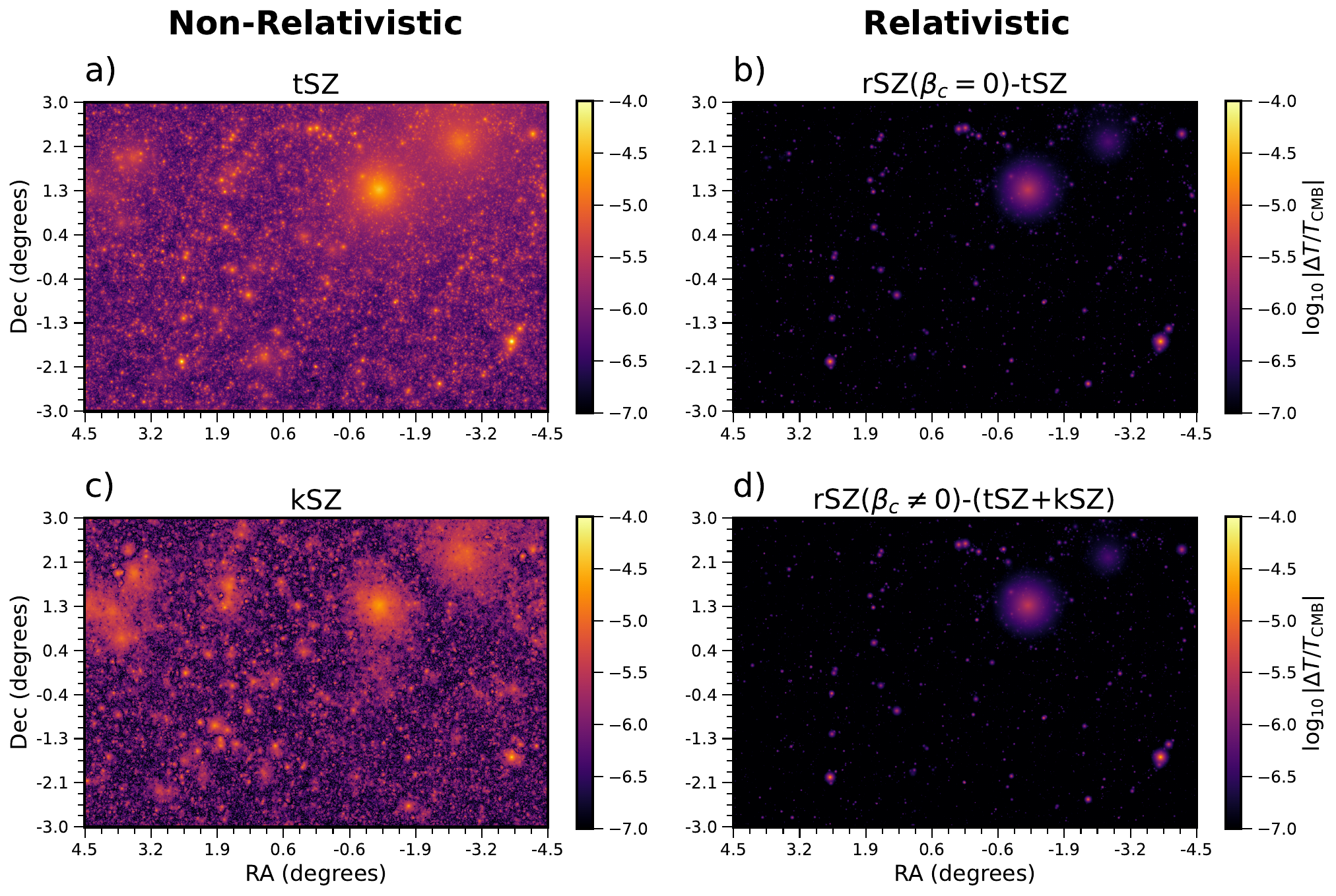}
\end{center}
\caption{Visualization of a painted map covering a small cutout region of the sky, spanning $6^\circ \times 9^\circ$ in area. The maps are constructed using a Clenshaw-Curtis variant workspace. Lettered panels illustrate the following: a) The thermal SZ (tSZ) signal, b) The difference between the tSZ and relativistic SZ (rSZ) signals, c) The kinematic SZ (kSZ) signal, d) The difference between the combined tSZ $+$ kSZ signals and the rSZ signal. Since visual differences between the tSZ/kSZ and rSZ signals are not expected, plotting the difference enables inspection of regions with the greatest variance. As anticipated, notable differences are observed primarily within the largest galaxy clusters. These maps were calculated using a frequency of 280 GHz.
}
\label{fig:rsz_cat}
\end{figure*}

Figure~\ref{fig:rsz_cat} shows the rSZ signal on a small cutout region of the simulated \textsc{websky} sky, $6^\circ \times 9^\circ$, in area. We choose to construct our maps using a CAR (Clenshaw-Curtis variant) pixelization. Panel a) shows the tSZ map of the sky, while panel b) shows the difference in the tSZ and relativistic corrections to the thermal SZ effect, obtained by evaluating \cref{eq:rSZ_full} at $\beta_c=0$. We do not expect to see visual differences on a log scale plot in the tSZ and rSZ signal so plotting the difference here allows us to inspect regions where we expect the greatest variance. One can visibly see that the relativistic corrections are most important for the highest mass (and therefore hottest) objects.  Panel c) shows the non-relativistic kSZ effect and Panel d) shows the difference between the full relativistic signal and the sum of the non-relativistic kSZ and tSZ effects. As can be seen the dominant relativistic effect are those from corrections to the thermal SZ effect; however, the ``cross terms" between the kSZ and tSZ term, $y\beta_c\mu  \mathcal{J}(\theta_e,x)$ in \cref{eq:rSZ_full}, are non-trivial and lead to small differences in Panels b) and d). These cross terms are examined more in Appendix \ref{sec:kineticCorrections}.

\cref{fig:rsz_fullsky} shows a full-sky rSZ map obtained from applying out method to the complete \textsc{websky} halo catalog. The catalog contains almost 1 billion dark matter halos and required three minutes per frequency map on 96 threads. We produce these maps at a range of frequencies and these will soon be available at \url{https://portal.nersc.gov/project/sobs/users/Radio_WebSky/rsz}. 

\begin{figure*}
\begin{center}
\includegraphics[width=0.98\textwidth]{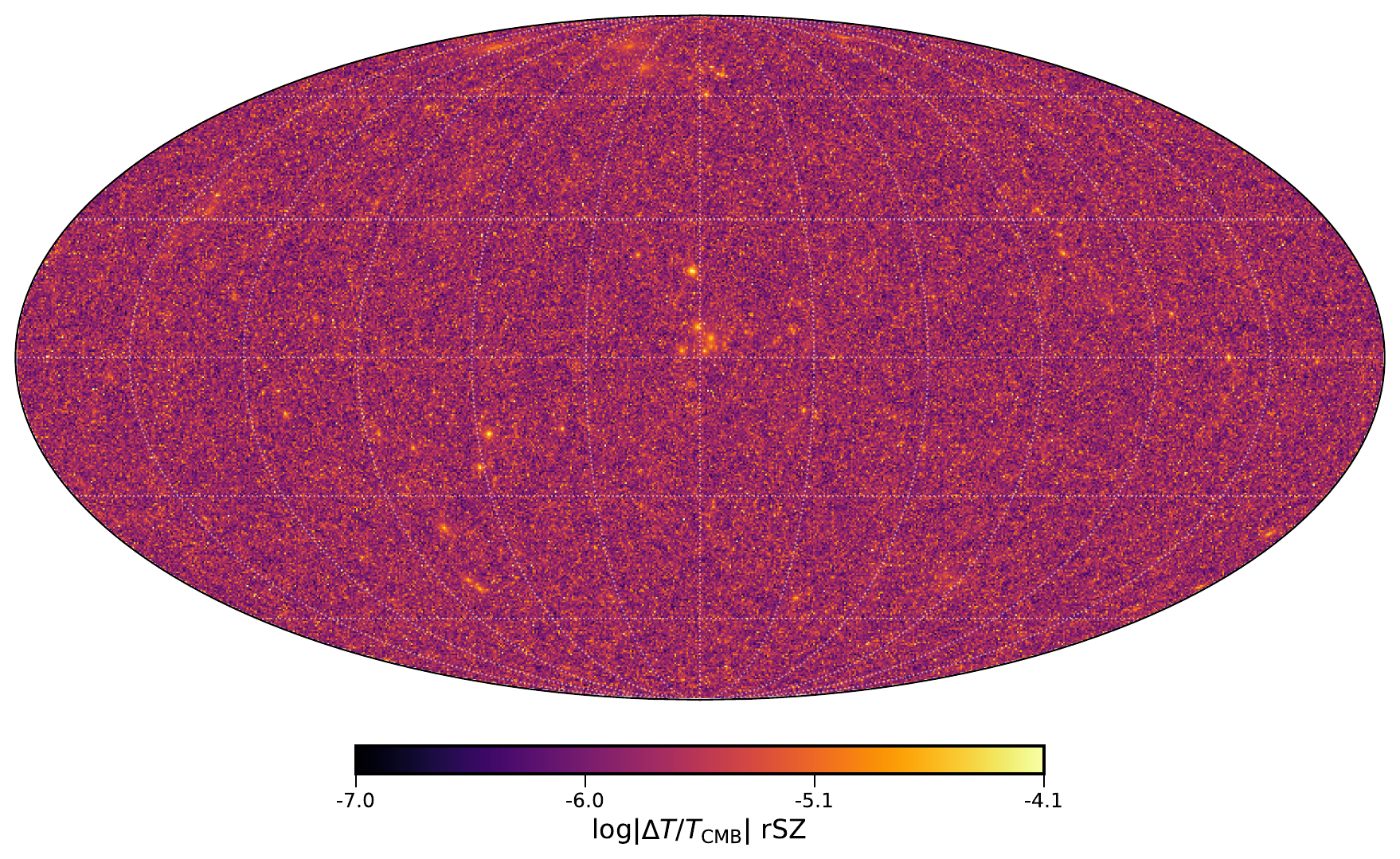}
\end{center}
\caption{Full sky simulated map of the rSZ traced as a temperature deviation from the CMB temperature. This figure was plotted using an \texttt{HEALPix} projection and was calculated at a frequency of 353 GHz.}
\label{fig:rsz_fullsky}
\end{figure*}

\section{From maps to physics}

To demonstrate the utility of these simulations, we perform a simulated analysis of the rSZ maps investigating what can be learnt about cluster temperatures with a Simons Observatory (SO)-like survey. In Sec. \ref{sec:forecast_configuration} we review the experimental configuration, in Sec. \ref{sec:forecast_methodology} the forecast methodology and present the results in Sec. \ref{sec:forecast_results_temps} and Sec. \ref{sec:forecast_results_pop}.

\subsection{Experimental Configuration for the forecasts}\label{sec:forecast_configuration}
For this forecast we generate simulated observations a Simons Observatory-like survey in combination with a simulated \textit{Planck} dataset. These mock observations include sky signals (including the CMB, extragalactic and Galactic processes) and instrumental effects (such as noise, the impact of the instruments finite resolution, i.e. the beam and  the observational footprint). We use the instrumental configuration given in Ref. \citep{2019JCAP...02..056A} for the SO-like experiment and in Ref. \citep{2020A&A...641A...1P} for \textit{Planck}. The SO-like survey has five frequency channels ranging from $27\,$GHz to $280\,$GHz and the deepest observations have a white-noise level of 8$\mu$K-arcmin. 
We use the \textit{Planck} channels between 30\,GHz and 545\,GHz with the deepest maps having white noise levels of 33$\,\mu$K-arcmin.  The sky signal uses either the non-relativistic tSZ from Ref. \cite{2020JCAP...10..012S} or the relativistic signal as implemented in this work. The other sky components are the \textsc{websky} maps of the kSZ, radio \citep{2022JCAP...08..029L}, CIB, lensed CMB, and the PYSM Galactic signals \citep{2021JOSS....6.3783Z}, using the ``d1" galactic dust, ``s1" synchrotron,``a1" anomalous microwave emission and '`f1"  free-free models \footnote{In this work we use more simple galactic foreground models, though consistent with observations. Dust is certainly an important concern when performing this analysis as it is dominant at high frequencies - where the rSZ signal is most different. However, our analysis setup here is not able to test the bias from Galactic dust. Our analysis should be thought of as a cross correlation: what is the rSZ like increment around objects in the galaxy catalog. In this work we use catalogs directly from the simulations without any contamination. Galactic dust features are uncorrelated with this true extragalactic catalog and so cannot bias our measurement. Note however that the noise from the galaxy is included in the analysis (and the galactic variance is well approximated with the ``d1'' and ``s1'' models as reported in \cite{2025ApJ...991...23T} ).}. We  apply beams with full-width-half-maxima (FWHM) given in   Ref. \citep{2020A&A...641A...1P}  and  Ref. \citep{2019JCAP...02..056A}. We assume a Delta function passband for each instrument. Finally we add instrumental isotropic noise based on the white-noise levels from \citep{2020A&A...641A...1P} and non-white atmospheric and instrumental noise as described in  Ref. \citep{2019JCAP...02..056A}. We consider an observational footprint covering $\sim30\%$ of the sky.

\subsection{Forecast Methodology}\label{sec:forecast_methodology}

The cluster temperatures are measured using the method proposed in Ref. \citep{2020MNRAS.494.5734R} and applied to \textit{Planck} \citep{Remazeilles_2024}, and Atacama Cosmology Telescope and \textit{Planck} data in Ref. \citep{coulton_2024b} and we refer the readers to these papers for further details. 

The key idea behind this approach is to represent the full relativistic thermal SZ effect as an expansion around a trial temperature, $\bar{T}_e$, as 
\begin{align}
&\delta I(\nu,\mathbf{n},T_e) = y(\mathbf{n})g(\nu,T_e) \nonumber \\
& =y(\mathbf{n})\left[ g(\nu,\bar{T_e})+\frac{\mathrm{d}g}{\mathrm{d} T_e} \left(T_e(\mathbf{n})-\bar{T}_e\right)+O\left( (T_e-\bar{T}_e)^2\right) \right] 
\end{align}	
Each term in this series has a different frequency dependence: the first term is $g(\nu,\bar{T}_e)$, the second is $g^{(1)}=\mathrm{d}g/\mathrm{d}T_e$ etc. Using component separation techniques \citep[see e.g.,][for a review]{Delabrouille_2007,leach2008} we can project multifrequency CMB observations into this basis and thereby make a map of the sky anisotropies with frequency dependence given by $g^{(1)}(\nu,T_e)$. What would we expect to see in such a map? Away from galaxy clusters, where $y(\mathbf{n})=0$, we expect to see noise. The signal seen at the location of a galaxy cluster will be $y(\mathbf{n})\left[T_e(\mathbf{n})-\bar{T}_e \right]$. Thus, if the trial temperature is lower (higher) than the cluster's temperature a negative (positive) signal will be seen. When the trial and cluster temperatures match no signal will be seen. In this analysis we iterate through a range of trial temperatures to find the trial temperature where the residual in the $g^{(1)}(\nu,\bar{T}_e)$ map is minimized, and this will correspond to the cluster temperature.

We implement this approach through the following steps: we use the Needlet Internal Linear Combination method (NILC) as first used in Ref. \cite{2009A&A...493..835D} and as implement in Ref. \citep{Coulton_2023} to isolate sky contributions with frequency dependence given by $g^{(1)}(\nu,\bar{T}_e)$. To minimize biases from the non-relativistic tSZ effect and the cosmic infrared background (CIB) we use the constrained ILC method \citep{Remazeilles_2011b}. This method explicitly mitigates biases from sky signals with a given frequency dependence. Note that since the CIB spectrum is not known perfectly and is likely to vary spatially, we model this with two components. First, a gray body with $T_d=10.7$ K and the spectral index $\beta_z=1.7$, which is a good approximation for the CIB spectral energy distribution (SED),  and a second term that captures deviations of the CIB from this assumed SED. Specifically we use the moment method to remove the first-order temperature derivative \citep[see][for more details]{Chluba_2017,Azzoni_2021,Coulton_2023,McCarthy_2023b}). CIB galaxies are distributed primarily in lower mass and higher redshift clusters \citep[see, e.g.,][for a review]{2018RvMP...90b5006K} and so we do not expect all clusters to be equally contaminated. 

In Appendix \ref{sec:noCIBdT} we demonstrate that lower mass clusters are more contaminated than massive clusters, to the extent that an additional constraint, which removes the derivative of the SED with respect to the spectral index, is needed in the NILC to mitigate CIB biases. Adding this constraint increases the noise in the measurement. We refer to this set of constraints as the 3 CIB  constraint NILC and the set above as the 2 CIB constraint NILC. Unless otherwise stated we report results with the 3 CIB constraint setup. The inputs to the NILC method are filtered to contain only scales with $500<\ell<4000$. Removing the largest scales, $\ell<500$, removes excess large-scale noise and removes modes larger than the cutouts (discussed below) that act as correlated noise in the stack. The small-scale cut is used as we need at least five high resolution observations, to satisfy the constraints in the ILC, and we do not have enough high resolution frequency observations to use scales beyond $\ell\sim 4000$.

We then extract 20$^\prime \,\times\,20^\prime$ cutouts around each object in our cluster catalog. The main catalog consists of all clusters with mass greater than $3\times 10^{14} M_\odot$ and so has $\sim 26,000$ objects after the cuts described below. This was chosen to match the expected number of clusters an SO-like experiment will detect.  We also include a very low mass sample with $1\times10^{14}M_\odot<M<3\times 10^{14}M_\odot$, similar to a set of clusters that could be optically detected by LSST\citep{2009arXiv0912.0201L} or Euclid \citep{2024arXiv240513491E}. To avoid biases from bright point sources (both radio and dusty star forming galaxies) we mask all radio sources with flux $>30$\,mJy at 90\,GHz and dusty galaxy with flux $>30$\,mJy at 220\,GHz. This is done in the input maps rather than applying a point source finder, as would be done on real data. Finally we apply a mask, based on the \textit{Planck} Galactic mask \citep{Planck_2018V}, to remove regions of the sky where the observations are dominated by Galactic signals. The cutouts can contain non-trivial levels of the kinetic SZ effect, which is not completely removed by the NILC method. The kSZ effect is larger than the typical size of the relativistic tSZ effect and thus could bias our measurement. We mitigate this effect by stacking the cutouts, using at least 500 objects per stack. The kSZ is mitigated as it has both a positive and negative sign (depending on whether the cluster is approaching or receding from the observer) and so will average to zero. 500 objects is sufficient that the residual kSZ is typically below $\sim 10\%$ of the relativistic tSZ effect \footnote{The primary CMB has the same spectral dependence as the kSZ but we are not worried about this contaminant for two reasons: firstly it has a very different scale dependence and secondly it is uncorrelated with the position of galaxy clusters so will only act as an easily-modellable source of noise.}. Note this averaging also removes the ``cross" tSZ-kSZ relativistic term. Finally, we compute 1D profiles, $d(r,\bar{T}_e)$, from the 2D stacks by azimuthally averaging the measurement.  
\begin{figure*}
  \begin{subfloat}[Non-relativistic  SZ Simulations\label{fig:sims_SZ}]{\includegraphics[width=.47\textwidth]{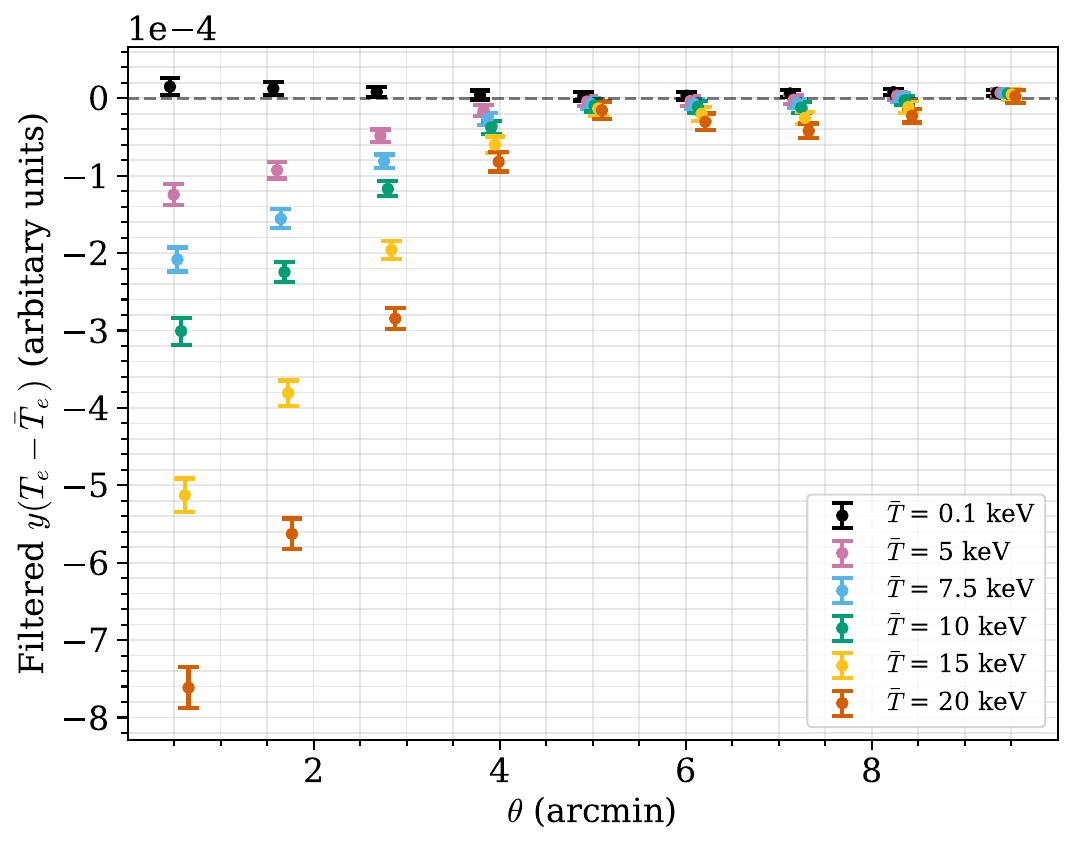} }
  \end{subfloat}
  \hfill
  \begin{subfloat}[Relativistic  SZ Simulations\label{fig:sims_rSZ}]{\includegraphics[width=.47\textwidth]{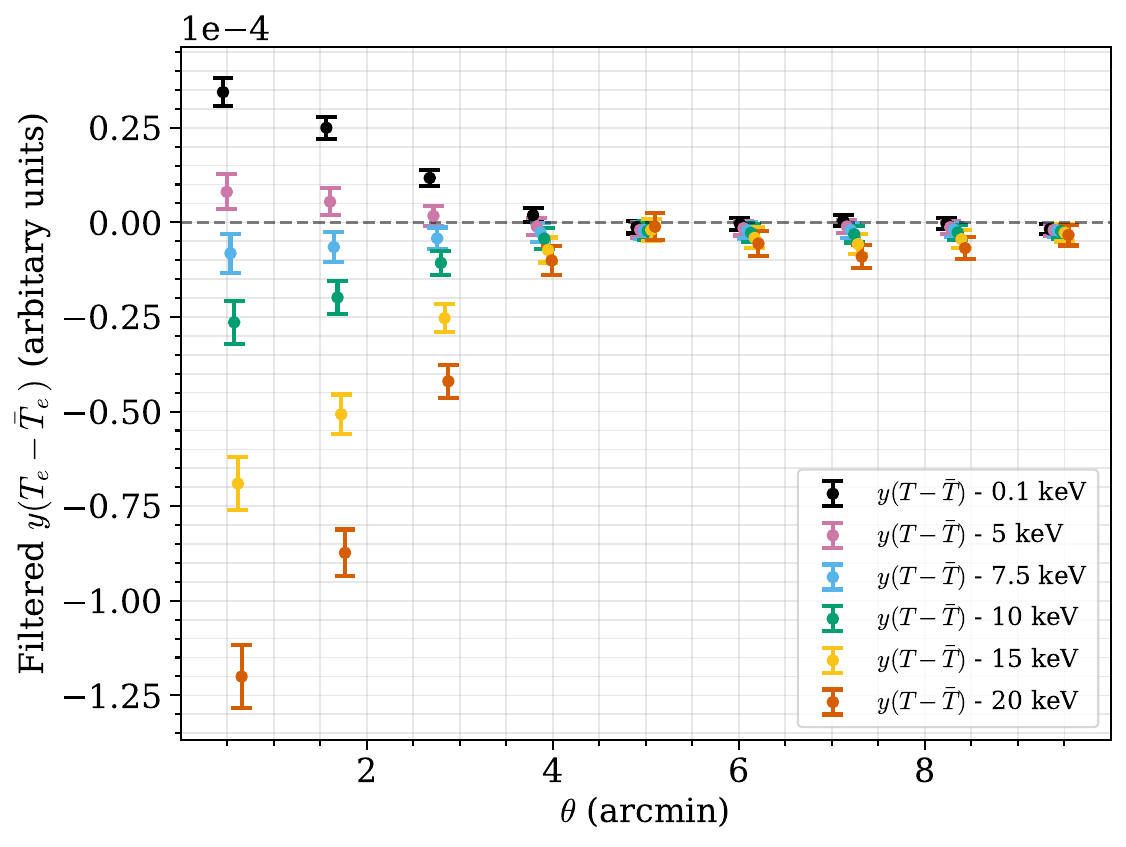} }
  \end{subfloat}
\caption{We apply the ``spectroscopic" method to a sample of clusters in simulated SO-like data. This method uses the relativistic corrections to measure the  the Compton-$y$ weighted difference between the true cluster signal and a trial temperature, $\bar{T}_e$. When the trial temperature is above (below) the true temperature, a negative (positive) residual is seen. A null is seen when the trial temperature matches the cluster temperature, as given by the rSZ spectral signature. The simulations in \cref{fig:sims_SZ} contain the non-relativistic thermal SZ effect and those in \cref{fig:sims_rSZ} use the relativistic tSZ simulations developed in this paper. The non-relativistic simulations favour zero temperature, as is expected as they lack the rSZ signal, whilst the measurements on the relativistic simulations show evidence for a non-zero temperature. }
\end{figure*}

\begin{figure*}
\begin{center}
\includegraphics[width=0.7\textwidth]{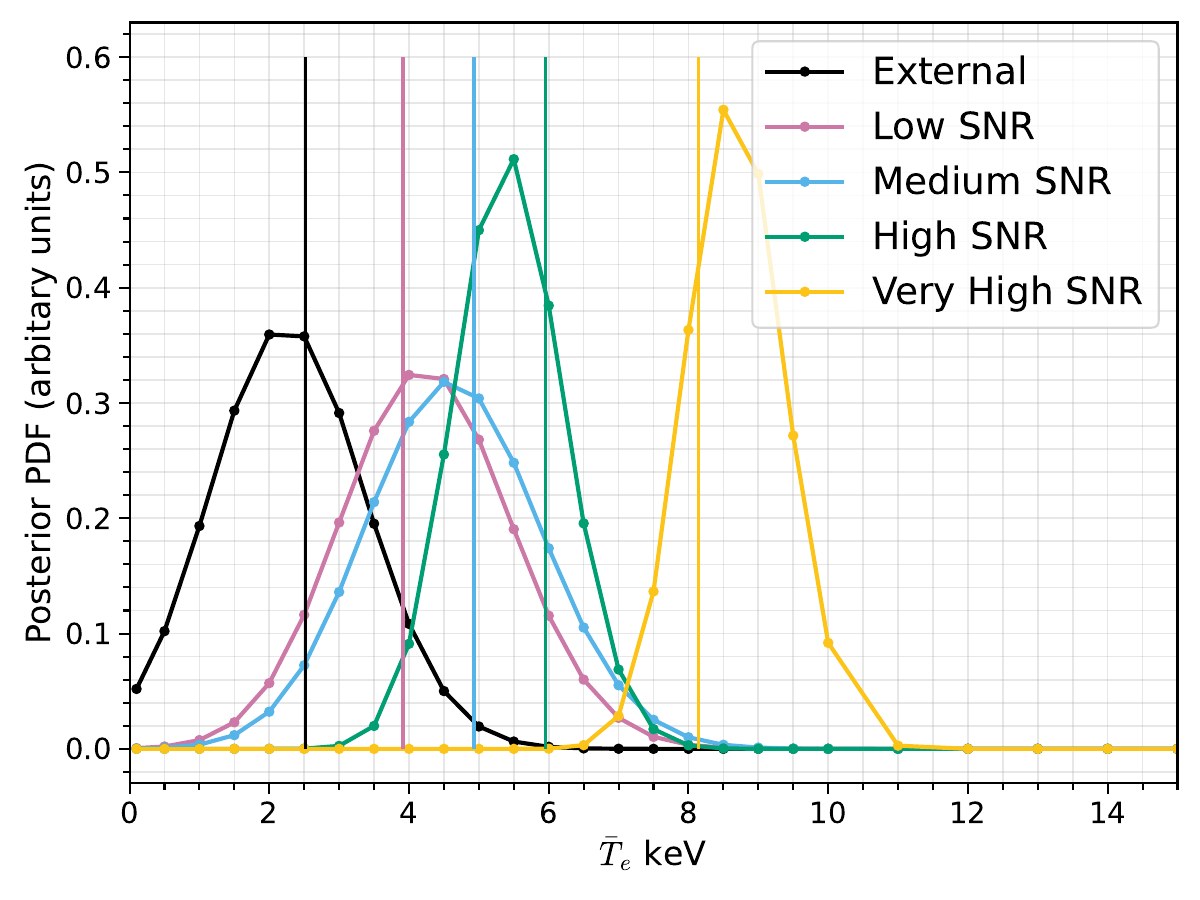}
\end{center}
\caption{Posteriors for the average temperature for five samples of clusters obtained from mock measurements with our relativistic SZ, SO-like simulations. The vertical lines denote the average input temperature. The samples, described in the text, are based on cluster mass. The external sample would not be detectable by the SZ effect, but could represent an optically selected sample.
This demonstrates that an SO-like experiment will be able to make precision, unbiased measurements of cluster temperatures.  \label{fig:temperature_posterior}}
\end{figure*}

We analyse these 1D profiles by computing the likelihood of any signal seen in the stack with a Gaussian likelihood with a prior as in Ref. \citep{coulton_2024b}. Explicitly, the log-posterior, $\ln \mathcal{P}$, has the following form
\begin{align}\label{eq:TePost}
\ln \mathcal{P}(T_e|d) \propto -\frac{1}{2} d(r,\bar{T}_e) C^{-1} d(r',\bar{T}_e),
\end{align}
where the $C$ covariance matrix of the 1D profiles and is computed from cutouts of the simulation far from any galaxy cluster.

A key ingredient in this likelihood is the covariance matrix. We estimate this from the mock observations by considering sky regions without any galaxy clusters. Specifically, we exactly repeat our procedure at randomly chosen positions on the map that are more than 6 arcmin from a galaxy cluster or bright point source. We use the same number of random objects as objects in the catalog. From this procedure we can compute the variance and covariance of the 1D profiles. This method is cross-checked by a second method based on negating the stacked signal. In this method we randomly select half the objects in the stack and negate them. We then recompute the stacked 1D profiles. The negation will remove the signal but not the noise. Repeated application of this procedure provides a second method of estimating the covariance matrix. The two methods provide consistent estimates of the covariance matrix.

\subsection{Constraints on cluster temperatures}\label{sec:forecast_results_temps}

To demonstrate how this works we apply this a set of the 1D profiles obtained from simulations containing the non-relativistic tSZ effect, \cref{fig:sims_SZ}, and to those containing the relativistic tSZ effect, \cref{fig:sims_rSZ}. This analysis uses the SO-like sample of clusters ($\sim 26,000$ clusters with $M>3\times 10^{14}M_\odot$). For all trial temperatures $\bar{T}_e>0.1\,$keV, the measurements on the non-relativistic simulation show a negative signal. This indicates that the trial temperature is higher than the true temperature (which is 0\,keV). For the relativistic simulations developed in this paper, we instead see a positive signal for low trial temperatures, indicating that the trial temperature is lower than the true temperature. For higher trial temperatures, a null between 5-7\,keV is seen and then a negative signal. This indicates that the mean temperature of these clusters is $\sim 6$\,keV.

Next we divide the cluster sample into five mass bins, defined as external: $1\times10^{14}M_\odot<M<3\times10^{14}M_\odot$, Low SNR: $3\times10^{14}M_\odot<M<5\times10^{14}M_\odot$, Medium SNR: $5\times10^{14}M_\odot<M<7\times10^{14}M_\odot$, High SNR: $7\times10^{14}M_\odot<M<10^{15}M_\odot$ and very high $M>10^{15}M_\odot$. For the largest two mass bins we use the 2 constraint NILC, as we find that these have less CIB contamination (see Appendix \ref{sec:noCIBdT}), whilst we use the 3 constraint NILC for the rest. We use \cref{eq:TePost} to obtain posteriors on the mean temperature of the clusters in the different stacks. \cref{fig:temperature_posterior} shows the measured temperature posterior for five cluster samples and the true temperatures used in the simulations. For this analysis we used the self-similar prediction for the temperature - mass relation \citep{2008ApJ...672..752W}. As expected, we measure a higher mean cluster temperature for the higher mass samples. Note that a zero cluster temperature is ruled out for each of the samples at 2.8, 3.8, 4., 7.6 and 14 $\sigma$. Thus an SO-like experiment will be able to detect the relativistic SZ effect at high significance.

\subsection{Constraining Cluster Sample Parameters}\label{sec:forecast_results_pop}

\begin{figure*}
\begin{center}
\includegraphics[width=0.98\textwidth]{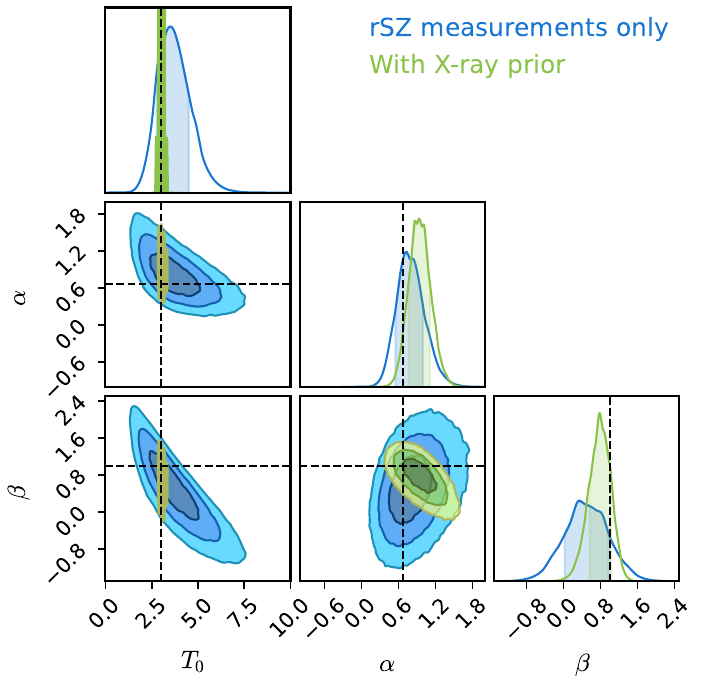}
\end{center}
\caption{We investigate how well an SO-like experiment can constrain the evolution of cluster temperature with mass and redshift based on the relation $T = T_0(M/M_\mathrm{pivot})^\alpha (1+z)^{\beta_z}$. The true values, given by the self-similar scaling relations, are shown in black. The green curves in show the results after imposing a tight constraint on $T_0$, which is a very crude approximation for the information accessible with X-ray observations. The green curve highlights synergies between SZ and X-ray cluster measurements. }
\label{fig:param_constraints}
\end{figure*}

In the previous section we used this method to measure the mean temperature for three different samples of clusters, thereby demonstrating that an SO-like experiment can detect this effect a high significance. In this section we explore whether such measurements can be used to place constraints on the properties of the cluster sample. Specifically, we examine whether these measurements can be used to constrain the mass and redshift dependence of the cluster temperatures. We assume a power-law form as
\begin{align}
T_e = T_0 \left(\frac{M_{200}}{M_{200,\mathrm{pivot}}} \right)^\alpha\left(1+z\right)^{\beta_z}
\end{align}
where $T_0$ is the modelled temperature of a $z=0$ cluster with mass equal to the pivot mass, $M_{200,\mathrm{pivot}}=6\times 10^{14}M_\odot$, $\alpha$ and $\beta_z$ govern the scaling of the cluster temperature with mass and redshift respectively. We will explore how much information can be extracted from the combined SO-like and \textit{Planck} dataset on the $T_0$, $\alpha$ and $\beta_z$ parameters. In addition to our CMB-based measurements, we will assume that masses and redshifts are available for the objects. We discuss this point further in the conclusions.

\begin{figure*}
  %\centering
  \begin{subfloat}[Reduced CIB mitigation (2 CIB constraint NILC) \label{fig:systematics_baseline}]{\includegraphics[width=.47\textwidth]{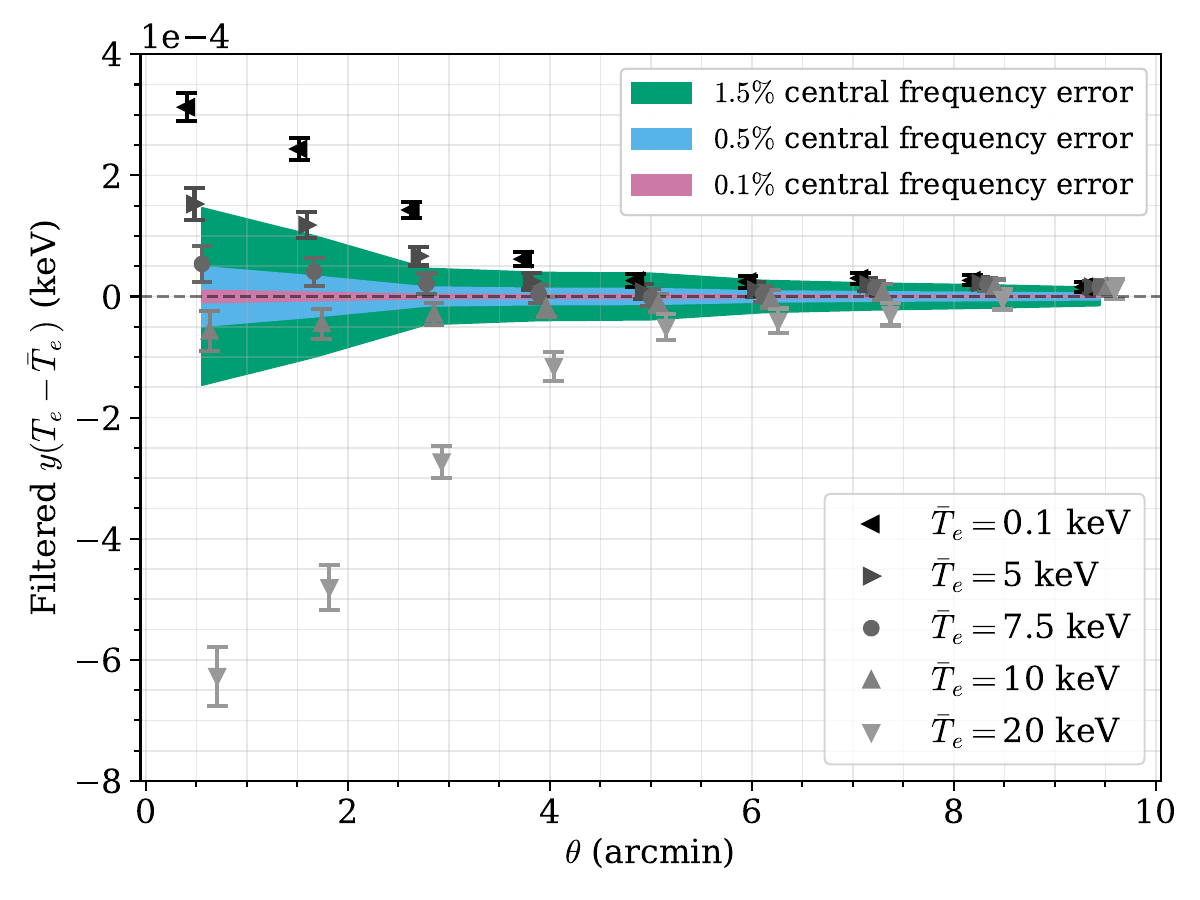} }
  \end{subfloat}
  \hfill
  \begin{subfloat}[Increased CIB mitigation (3 CIB constraint NILC)\label{fig:systematics_enhanced}]{\includegraphics[width=.47\textwidth]{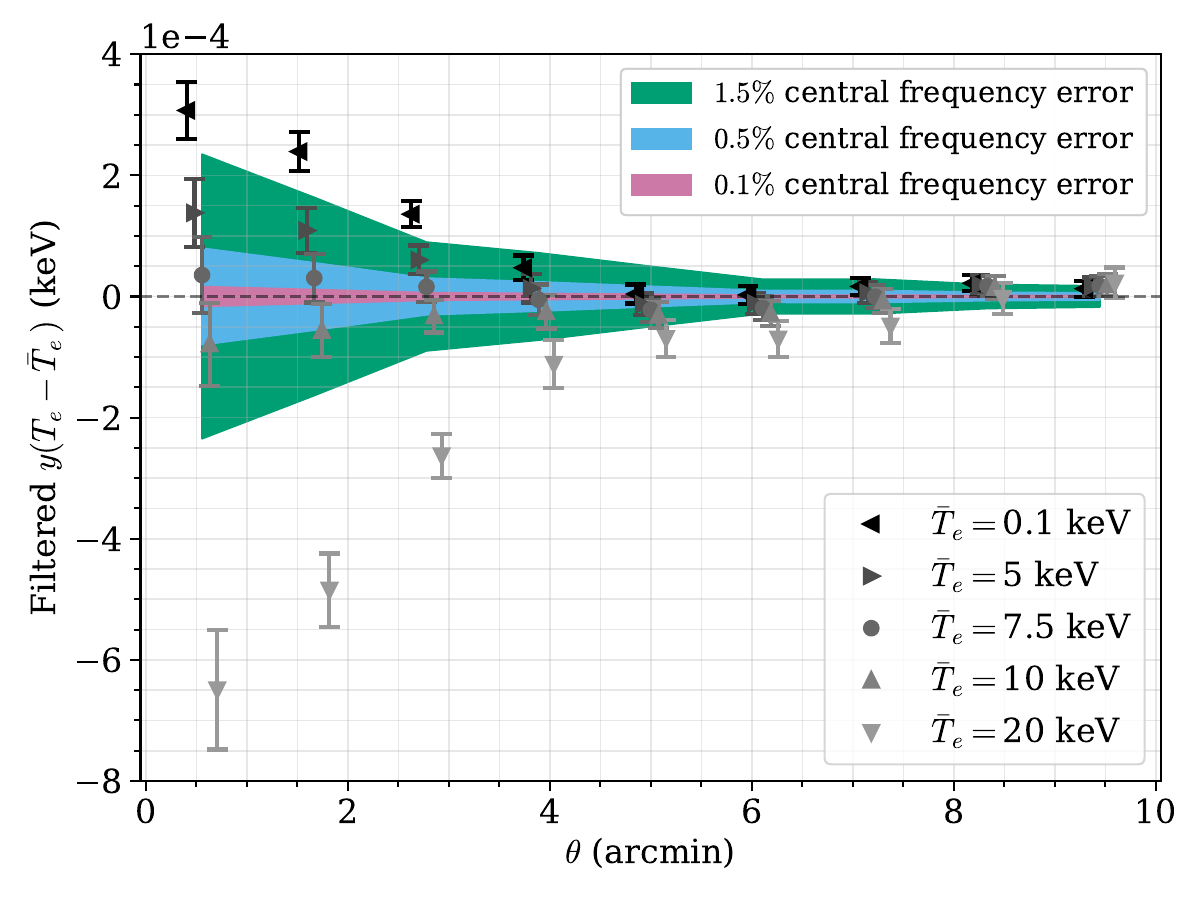} }
  \end{subfloat}
\caption{Measuring the rSZ effect requires a highly precise characterization of the instrument, especially the passbands. We examine the impact of a 0.1\%, 0.5\%, and 1.5\% uncertainty in the central frequency of passbands on these measurements. We show the $1\sigma$ uncertainty band compared to our measurements of the high mass cluster sample ($M>10^{15}M_\odot$) for the 2 constraint NILC (\cref{fig:systematics_baseline}) and for the 3 constraint NILC (\cref{fig:systematics_enhanced}). Passband systematics are measured for the sample with $\bar{T}_e=5\,$keV and set around the x-axis to aid interpretation. These systematic errors will dominate the statistical uncertainty, unless the error on the central frequency can be reduced below $0.5\%$. We also see that the passband uncertainties more strongly impact the 3 CIB constraint measurement.\label{fig:systematics}}
\end{figure*}

To do this we create stacked 1D profiles, as described in Sec. \ref{sec:forecast_methodology}, for many samples of clusters binned by mass and redshift. First we create 20 mass bins: ten low mass bins, which are linearly spaced between $1.25\times 10^{14} M_\odot$ and $3.8\times 10^{14} M_\odot$, and ten high mass bins. The latter bins are chosen by creating a bin from the 2000 highest mass objects that are not yet in a bin, and then repeating this procedure. This joint approach is used to balance the fact that there are many more low-mass objects than higher mass objects with the need to have a broad mass range to efficiently constrain the mass evolution parameters. Next each bin is divided into 4 redshift bins, based on the clusters' redshift. For each bin we then compute the posterior for the mean cluster temperature, $p_i(T_e | \left\{(M_i,z_i)\right\})$, where $\left\{(M_i,z_i)\right\}$ denotes the redshifts and mass for each object in the stack. Taking all bins together, we have a 2D probability surface for the cluster temperature as a function of mass and redshift, which we can use to constrain the parameters of the temperature-mass-redshift relation. We assume a uniform prior on $T_0$ between $0$\,keV and $35$\,keV and uniform priors between $-3$ and $3$ for $\alpha$ and $\beta_z$. 

In \cref{fig:param_constraints} we show the resulting constrains on the temperature-mass-redshift parameters. The results show that it will be possible to constrain all three of these parameters as there is a significant reduction of the prior volume space. The best constrained parameter is the mass evolution parameter, which is constrained to be $0.74_{-0.19}^{+0.25}$, cf the truth of 2/3. However, it can be clearly seen that there are degeneracies between all the parameters. The strongest degeneracy is found between the normalization temperature $T_0$ and the redshift evolution parameter $\beta_z$. This arises as the redshift evolution is not well constrained by this simulated sample.

One potential path to remove the degeneracy between pivot temperature and the redshift evolution parameter is to exploit the complementarity of other observational probes. X-ray measurements provide an alternative path to measuring cluster temperatures, though for large samples current experiments are limited to lower redshifts. A simple demonstration of the complementary of this is shown in \cref{fig:param_constraints} where we add a tight Gaussian prior on $T_0$ with width $0.1$ FWHM. This prior very crudely mimics having precision lower redshift information about a high mass cluster sample. In this case we are able to dramatically improve the constraint on the redshift evolution of cluster temperatures.  Conversely these observations will allow interesting comparisons between the thermodynamics inferred by SZ and X-ray.

\subsection{The importance of instrumental systematic errors}\label{sec:systematics}

The relativistic corrections are small ($\sim$ few percent) corrections to the SZ effects. Detecting these effects thus requires a precise understanding of the instrument. This was explicitly shown in Ref. \citep{coulton_2024b}, where it was found that instrumental systematic uncertainties for ACT limit the ability to measure the rSZ effect. Specifically, they found that uncertainties in the central frequency of each observation band limited the measurement. In this section we explore how uncertainties in passband measurements will impact future rSZ measurements.

We used the procedure described in Ref. \citep{Coulton_2023} to investigate the impact of systematic errors. In this approach we sample a central frequency shift for each observation frequency channel, apply this to the observation bands, and then reanalyze the data. This is repeated, and the resulting distribution shows the impact of instrumental uncertainties on the measurements. We sample the frequency shifts from a Gaussian distribution centered on zero with three different standard deviations: $1.5\%$, $0.5\%$ and $0.1\%$ of the central observation frequency. These roughly correspond to the uncertainty on ACT passbands from Fourier Transform Spectrometer (FTS) measurements, the expected uncertainty for FTS measurements on a SO-like instrument and a goal level.

In \cref{fig:systematics} we compare the statistical errors on the measurements of the high mass sample (the best measured sample) to the statistical errors. We see that the systematic errors dominate over the statistical errors for the $1.5\%$ and $0.5\%$. Thus, to avoid being systematics limited, passband central frequencies need to be better than $0.5\%$. It can also be seen that more aggressive CIB cleaning methods (e.g., using two CIB moments instead of one) increase the impact of systematic errors. This means that systematic errors will be more problematic for lower mass cluster samples, for which CIB contamination is worse.

\section{\label{sec:conclusion}Conclusion}

The SZ effect provides a unique way to map the large-scale structure of the universe as traced by ionized gas. When the temperature or velocity of the electrons in the plasma becomes non-negligible, observations of the rSZ can provide additional information about the detailed temperature and velocity structure of the ICM. In this work, we have developed a computational toolkit, integrated into the \texttt{XGPaint} package, to evaluate rSZ signals and paint them onto lightcone halo catalogs. We applied this code to the \textsc{Websky} simulation suite to generate full-sky rSZ simulations.   

A key ingredient in simulating the rSZ effect is the assumed electron temperature. We test several mass-temperature scaling relations, which were obtained from analyzing different hydrodynamical simulations. We find that, in terms of the angular distribution of the rSZ signal, there are significant differences in the predicted cluster signal depending on how AGN feedback is implemented (see \cref{fig:t_scale}). This demonstrates that rSZ measurements can be used to probe the thermal history of gas in galaxy groups and clusters.

Our simulated rSZ maps have implications for many upcoming sub-millimeter instruments, including SO and FYST. SO's planned frequency bands align nicely with the important regions of the rSZ spectrum (see \cref{fig:rsz_temp}). This will enable direct comparisons of simulated and observed rSZ maps to evaluate the compatibility of theory and observation. Our new mock rSZ observations will help us understand the capabilities for future instruments.  Further, these mocks are correlated with other large-scale structure tracers and cross-correlations with other large-scale sky tracers, as simulated in the \textsc{websky} extragalactic simulations, can provide more insight into the formation and evolution of the large-scale structure.

As a simple demonstration of the utility of these maps, we performed a forecast to investigate the sensitivity of an SO-like experiment, in combination with the \textit{Planck} satellite, to the relativistic SZ effect. This analysis focused on using the rSZ to probe the temperature of clusters. We found that an SO-like experiment will be able to make precision measurements of the mean cluster temperature for stacks of galaxy clusters. For example, we forecast that the temperature of $\sim26000$ clusters can be measured in at least four mass bins with high SNR. Further, we showed that the population properties of the sample can be effectively constrained by considering how well the parameters of a mass - redshift - cluster temperature relation can be measured (see \cref{fig:param_constraints}). This analysis found a significant degeneracy between the pivot temperature and the redshift evolution parameters arising from the more limited number of very high redshift clusters. 

Whilst this analysis is encouraging for future studies of the relativistic SZ effects, it has several limitations. It largely ignores the impact of observational systematics, and these are expected to have a significant impact on this measurement. To underscore this we investigated one instrumental effect: how different levels of uncertainty in the central frequency of the passbands impact our forecast. We find that systematic errors dominate the statistical errors unless the central frequencies are known to a level better than $0.5\%$. To reach the statistical precision of next generation experiments, the instrument's beams, calibration and transfer functions will need to be equally precisely known. Secondly, in our analysis we assumed that the redshift and mass of each cluster were known perfectly. Whilst there are established methods for measuring cluster redshifts with follow- ups\citep[see e.g.,][and references therein]{2021ApJS..253....3H,2024OJAp....7E..13B}, it is still difficult to measure cluster masses at high precision. Optical and CMB weak lensing measurements provide one path for future observations \citep[see e.g.,][]{2019MNRAS.489..401Z,2023arXiv231012213B}. An alternative would be to constrain the relationship between other observational parameters, such as the integrated Compton-y signal, and cluster temperature. Work by Ref. \citep{2022MNRAS.517.5303L} has already shown that there are interesting physical correlations between such observables.

More abstractly this forecast demonstrates one anticipated use of these simulations - to validate analysis pipelines and to explore methods of measuring the relativistic SZ effect. This analysis used the spectroscopic method, proposed by Ref. \citep{2020MNRAS.494.5734R}, in combination with a binning of the galaxy cluster catalog. This method is certainly not optimal as the binning throws away information. Thus, it would be interesting to see if more powerful methods can be developed. A second challenge is the issue of contamination from other sky signals. In this work we used the constrained ILC method to mitigate biases from the CIB and a masking approach for radio galaxies. These are crude methods, which produced unbiased results in this study, and it would be very interesting to explore alternative methods.

\section*{Acknowledgments}
The authors are grateful to Jens Chluba for comments on the draft. LK thanks the CITA SURF program for support during the initial stages of this project. 
\appendix

\section{Beyond corrections to the thermal SZ effect \label{sec:kineticCorrections}}
\begin{figure*}
\begin{center}
\includegraphics[width=0.98\textwidth]{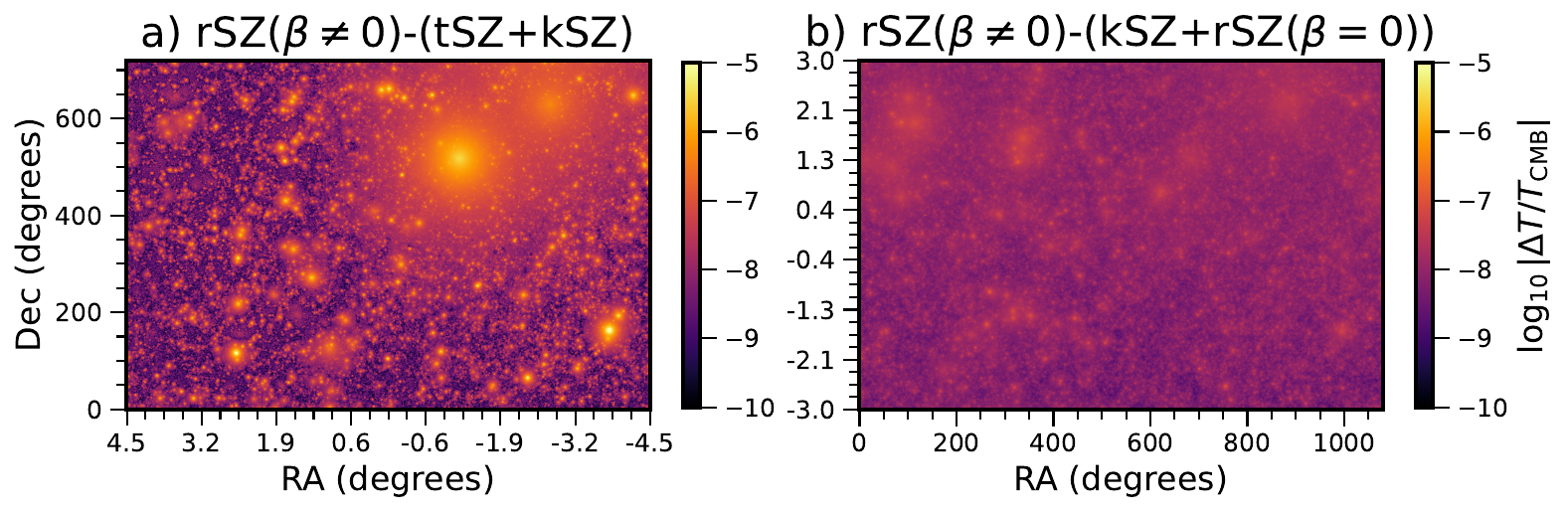}
\end{center}
\caption{This plot examines the size of the velocity dependent rSZ terms by removing the non-relativistic kSZ and relativistic thermal SZ effect from the total rSZ signal (panel b). To aid interpretation we also show the size of total rSZ corrections in panel a) - note that this is the same as panel d) from \cref{fig:rsz_cat} with a different colorbar. As expected the velocity terms are significantly smaller, though not completely negligible. }
\label{fig:rsz_cat_kszTerms}
\end{figure*}

As demonstrated in \cref{fig:rsz_cat}, the dominant relativistic correction is the correction to the thermal SZ effect, i.e., higher order temperature corrections at $\beta_c=0$; however the velocity terms and the cross terms (thermal corrections to the kSZ effect and vice-versa) are non-negligible and of cosmological interest (see e.g. \cite{2020PhRvL.125k1301C}). We isolate these terms in \cref{fig:rsz_cat_kszTerms}. These terms are typically an order of magnitude smaller than the dominant rSZ correction.  

\section{The challenges of removing the CIB\label{sec:noCIBdT}}

\begin{figure*}
  \begin{subfloat}[High mass sample\label{fig:noDT_highMass}]{\includegraphics[width=.47\textwidth]{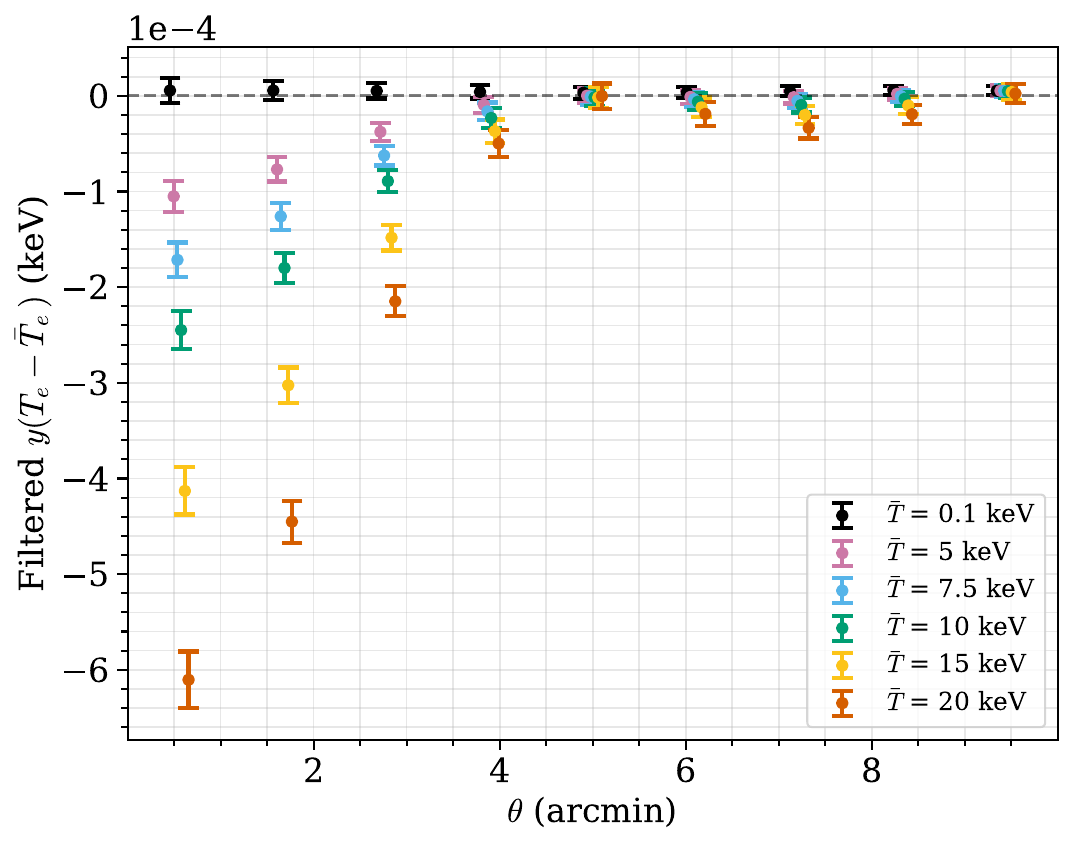} }
  \end{subfloat}
  \hfill
  \begin{subfloat}[Low mass sample\label{fig:noDT_lowMass}]{\includegraphics[width=.47\textwidth]{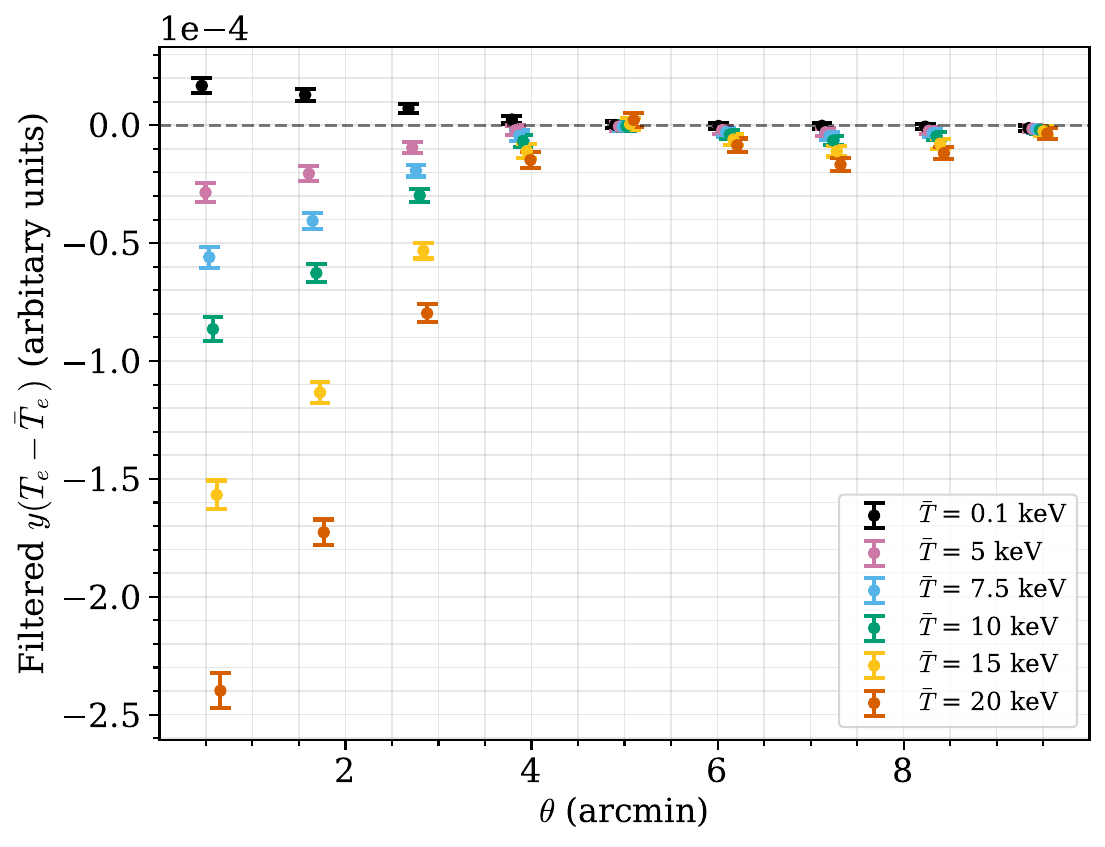} }
  \end{subfloat}
\caption{An assessment of potential contamination from the CIB. We perform an analysis of the non-relativistic simulations without the third CIB mitigation method (i.e. in the NILC we deproject the CIB and the temperature derivative of the CIB SED components only). \cref{fig:noDT_highMass} shows  the analysis of the high mass sample ($M>10^{15}\,M_\odot$) and \cref{fig:noDT_lowMass} shows a lower mass sample ($2\times 10^{14}\,M_\odot<M<5\times 10^{14}\,M_\odot$). The non-relativistic simulations lack any rSZ signal, so the temperatures should be consistent with zero. Deviations from this indicate a bias. The high mass sample is robust; however, the lower mass sample is biased by the CIB. This bias is removed by deprojecting an additional component in the ILC, with the results shown in \cref{fig:sims_SZ}. Note that the sensitivity to the CIB depends on the experimental configuration and measurement precision, and the assumptions on the CIB.\label{fig:noDT}}
\end{figure*}

In the NILC we primarily use three constraints to mitigate CIB contamination. This results in a large noise penalty. It is worth exploring whether less aggressive mitigation approaches can obtain unbiased inferences. In \cref{fig:noDT}, we apply our method to non-relativistic simulations and remove one of the NILC constraints. We see evidence for non-zero temperature, even though the simulation is non-relativistic, i.e. the analysis is biased. For higher mass samples, smaller biases are seen. This suggests that for the CIB model in Websky, all the conditions are necessary.  Modelling the CIB is very challenging, and different simulations will find different conclusions. Thus, it is important to test for CIB contamination in the analyses of experimental data to ensure there are no significant biases.

\section{Mass Uncertainties}
\begin{figure*}
\begin{center}
\includegraphics[width=0.98\textwidth]{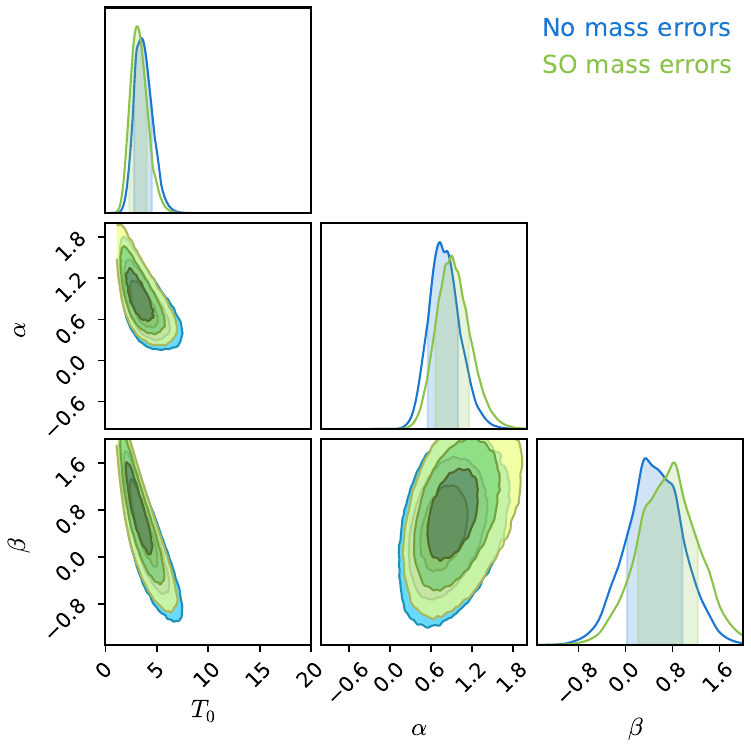}
\end{center}
\caption{An estimation of the impact of cluster mass errors on our inferred scaling relations. We compare the constraints on our power law model,  $T = T_0(M/M_\mathrm{pivot})^\alpha (1+z)^{\beta_z}$, from a data vector where the mass are known perfectly (blue) to the case with mass uncertainties consistent with CMB lensing mass estimates (green). The resulting contours are broadly consistent showing that this is expected to be a subdominant source of bias. }
\label{fig:impactOfMassErrors}
\end{figure*}

In this work we assumed perfect knowledge of the cluster masses. However, cluster mass estimation is difficult and is the subject of many investigations \citep[e.g.][]{2020A&ARv..28....7U,2019MNRAS.489..401Z,2024PhRvD.110h3510B}. To test the importance of this we perform a simple test to assess the robustness of our results to mass uncertainties. We apply a gaussian scatter to the masses of the samples. We use $\sigma(M)/M = 0.04$ for a sample of 1000 clusters and scale by $\sqrt(N)$ to match our sample sizes. This fractional error is broadly consistent with the expected CMB lensing mass calibration error, e.g.
\cite{2019JCAP...02..056A}. We then repeat the analysis with these shifted bins. The result is shown in Fig. \ref{fig:impactOfMassErrors}  and the contours are fairly consistent - indicating that mass errors, at the expected level with CMB-lensing calibration, are likely not a major issue. We defer a more detailed investigation to future work.

% Bibliography

%% [A] Recommended: using JHEP.bst file
\bibliographystyle{JHEP.bst}
\bibliography{rsz}

%\begin{thebibliography}{99}
%\end{thebibliography}
\end{document}